\documentclass[%
 reprint,
floatfix,
superscriptaddress,
showpacs,preprintnumbers,
 amsmath,amssymb,
 aps,
pra,
]{revtex4-2}
\DeclareUnicodeCharacter{2212}{-}
\usepackage{xcolor}
\usepackage{times}
\usepackage{amssymb}
\usepackage{dsfont}
\usepackage{graphicx}% Include figure files
\usepackage{dcolumn}% Align table columns on decimal point
\usepackage{bm}% bold math
\usepackage{epstopdf}
\usepackage[colorlinks]{hyperref}% add hypertext capabilities
\hypersetup{colorlinks = true,
linkcolor = blue,
filecolor = blue,
urlcolor = blue,
citecolor = blue}

\begin{document}

\title{\textbf{Probing quasiparticle excitations in a doped Mott insulator via Friedel oscillations}}
\author{Anurag Banerjee}
\thanks{These two authors contributed equally}
\affiliation{Universit\'{e} Paris-Saclay, Institut de Physique Th\'eorique,  CEA, CNRS, F-91191 Gif-sur-Yvette, France}

\author{Emile Pangburn}
\thanks{These two authors contributed equally}
\affiliation{Universit\'{e} Paris-Saclay, Institut de Physique Th\'eorique,  CEA, CNRS, F-91191 Gif-sur-Yvette, France}

\author{Catherine P\'epin}
\affiliation{Universit\'{e} Paris-Saclay, Institut de Physique Th\'eorique,  CEA, CNRS, F-91191 Gif-sur-Yvette, France}

\author{Cristina Bena}
\affiliation{Universit\'{e} Paris-Saclay, Institut de Physique Th\'eorique,  CEA, CNRS, F-91191 Gif-sur-Yvette, France}

\begin{abstract}
%Impurities offer a powerful probe into the nature of low-energy excitations in strongly correlated systems. In weakly interacting metals, impurity-induced phenomena such as quasiparticle interference and Friedel oscillations reveal key features of the Fermi surface. However, the nature of analogous responses in strongly correlated systems remains less understood. 
In this work, we investigate impurity-induced Friedel oscillations in the doped two-dimensional Hubbard model, focusing on the role of holon and doublon excitations. We show that weak impurities, due to the non-fermionic nature of the underlying quasiparticles, induce Friedel oscillations whose behavior is consistent with an effective non-interacting theory for these quasiparticles, and whose wavevector reflects the violation of Luttinger's theorem. At larger impurity strength, the system transitions to a phase-separated state composed of coexisting Mott-insulating (half-filled) and hole-rich regions. Within the composite operator framework, this phase separation arises from a competition between the kinetic energy of holons and the tendency to form tightly bound holon-doublon pairs. Our results offer new insights into the nature of charge carriers and the emergent electronic phases in the doped Mott regime.
\end{abstract} 

\maketitle

\section{Introduction}
In non-interacting systems, quasiparticle interference (QPI) patterns~\cite{wang2003quasiparticle,balatsky2006impurity}, which arise from electron scattering off impurities, are a well-established tool for probing Fermi surface properties via surface scanning tunneling microscopy (STM) experiments. Furthermore, by analyzing the scattering of low-energy degrees of freedom off impurities, insights can be gained about the nature of the scatterers. While in a weakly correlated system, the scatterers of an electronic system are dressed electrons, their nature can change drastically with strong interactions. A striking example is the one-dimensional Hubbard model, where electrons fractionalize into independent spin and charge degrees of freedom~\cite{mudry1994separation}.  Recent work has demonstrated that Friedel oscillations can reveal the crossover between a spin-$1/2$ Luttinger liquid and a spinless chargon liquid in the one-dimensional Hubbard model~\cite{bohler2025probing,zeng2024quasiparticle}.

The nature of spin and charge carriers in the doped two-dimensional Hubbard model remains an open and actively debated question~\cite{arovas2022hubbard}. Investigating how impurities affect low-energy properties can thus provide crucial insights into the emergent quasiparticles in these systems. This strategy has been influential in the cuprate pseudogap phase, where impurity scattering signatures through the octet model have improved our understanding of the underlying electronic structure~\cite{zeng2024quasiparticle}. Even at half-filling, the nature of excitations remains nontrivial, with theoretical proposals suggesting the existence of exotic chargeless spinon Fermi surfaces~\cite{mross2011charge}, intimately connected to zeros of the Green's function~\cite{wagner2024edge}. Such states have given rise to Friedel oscillations in certain systems described by exactly solvable models~\cite{zhao2023friedel}.

This work adopts a complementary approach by focusing on the local charge excitations, holons and doublons, near a Mott insulating state. These excitations are naturally described within the framework of Hubbard operators~\cite{hubbard1963electron,hubbard1964electron}, which obey non-canonical anticommutation relations. Consequently, even if holons and doublons behave like weakly interacting quasiparticles, the resulting metallic state exhibits non-Fermi liquid characteristics. While double-occupancies are often deemed irrelevant in the large-$U$ limit, their role cannot be dismissed outright. Simply integrating out doublons may obscure important physics associated with ultraviolet-infrared (UV-IR) mixing in the spectral function~\cite{phillips2010colloquium}, sometimes referred to as a "color change" phenomenon~\cite{hirsch1992superconductors,molegraaf2002superconductivity}.  Moreover, the metal-insulator transition at half-filling driven by increasing Coulomb repulsion can be fruitfully understood as a binding transition of holons and doublons~\cite{castellani1979new,kaplan1982close,yokoyama1987variational,imada1998metal,capello2005variational,choy2008exact,prelovvsek2015holon}. This picture provides a unifying framework for interpreting the nature of excitations both near and away from half-filling in strongly correlated systems.
%In this work, we take a different approach by focusing on the fundamental local charge excitationsholons and doublons in the proximity to a Mott state. These emergent excitations are described by Hubbard operators~\cite{hubbard1963electron,hubbard1964electron}, which obey non-canonical anticommutation relations. This implies that if these excitations behave as weakly interacting quasiparticles, the resulting metallic state deviates from a conventional Fermi liquid, exhibiting non-Fermi liquid behavior. Although doublons may appear irrelevant in the large $U$ limit, simply integrating them out can miss the physics arising from UV-IR mixing~\cite{phillips2010colloquium}, also referred to as color change~\cite{hirsch1992superconductors,molegraaf2002superconductivity}, in the electronic spectral function. Moreover, the metal-insulator transition at half-filling, driven by increasing Coulomb interaction, can be understood in terms of the binding of holons and doublons~\cite{castellani1979new,kaplan1982close,yokoyama1987variational,imada1998metal,capello2005variational,choy2008exact,prelovvsek2015holon}.

We investigate the role of impurities in a doped Mott insulator, focusing on two distinct regimes characterized by the strength of the impurity potential. In the weak-impurity regime, we observe the emergence of Friedel oscillations (FO), consistent with prior findings in correlated metals and doped Mott insulators~\cite{ziegler1998friedel,soffing2009wigner,chatterjee2019real,zhao2023friedel,bohler2025probing}. The wavevector associated with these oscillations, as extracted from a Fourier transform of the local density of states (known also as a quasiparticle interference pattern or QPI), reflects a fundamental violation of Luttinger's theorem, as captured within the Hubbard operator framework. Importantly, this violation arises not from the specifics of any approximation scheme, but as a direct consequence of the non-canonical (non-fermionic) algebra obeyed by the holon and doublon operators. We show that the dispersion of the QPI patterns is consistent with an effective non-interactive model for the holon and doublon excitations.

In the strong-impurity regime, non-perturbative effects drive the system into a phase-separated state characterized by spatial coexistence of half-filled Mott insulating regions and hole-rich metallic domains. This emergent phase bears strong resemblance to phase-separated ground states proposed in earlier studies of the Hubbard~\cite{visscher1974phase,hellberg1997phase,aichhorn2006variational,macridin2006phase,ma2021interacting,seufert2021breakdown,riegler2023interplay} and $t$-$J$ models~\cite{schrieffer1988spin,emery1990phase,marder1990phase,moreo1991phase,putikka1992aspects,emery1993frustrated,eder1994spin}. Within the holon-doublon representation, the phase-separated state can be understood as the outcome of a competition between the kinetic energy gained by delocalized holes and the localization tendency arising from tightly bound holon-doublon pairs~\cite{hansen2022doping,kapetanovic2024charge}.

The paper is organized as follows. Section~\ref{Sec:Method} provides a brief overview of the composite operator formalism employed to study the Hubbard model in the strong-coupling limit ($U \gg t$). Section~\ref{Sec:LowImp} presents self-consistent results in the weak impurity regime, emphasizing the emergence of anomalous Friedel oscillations as a manifestation of Luttinger theorem violation. In Section~\ref{Sec:LargeImp}, we explore the onset of phase separation induced by strong impurities, identifying its origin in the interplay between holon delocalization and the formation of tightly bound holon-doublon pairs. We discuss the implications of our findings and conclude in Section~\ref{Sec:Discussions}. For completeness, Appendix~\ref{App:COMDetails} outlines the key steps of the equation-of-motion method used in our analysis.

\section{Model and Method\label{Sec:Method}}
\subsection{Model}
We work with the repulsive Hubbard Hamiltonian, which is given by
\begin{align}
\mathcal{H}=&-\sum\limits_{\langle ij\rangle,\sigma}\big(t_{ij}c^\dagger_{\sigma}(i)c_{\sigma}(j)+h.c.\big) \\
\nonumber&+U\sum\limits_i \hat{n}_{\uparrow}(i) \hat{n}_{\downarrow}(i)-\mu\sum_{i,\sigma} \hat{n}_{\sigma}(i).&
\label{eq:eqHubbard}
\end{align}
Here, $c_{\sigma}(i)$ $(c^\dagger_{\sigma}(i))$ annihilates (creates) an electron at site $i$ with spin $\sigma$, where $\sigma=\uparrow,\downarrow$ for spin-1/2 electrons. The first term represents hopping between neighboring sites $i$ and $j$, typically nearest neighbor hopping denoted by $t_{ij}=t$ if $i$ and $j$ are nearest neighbors; otherwise, zero. We set $t=1$ in this work and all our energy scales are in the units of $t$. The second term accounts for on-site repulsion between electrons, characterized by the strength $U>0$. The number operator is defined as $\hat{n}_{\sigma}(i)= c^\dagger_{\sigma}(i) c_{\sigma}(i)$, and the local electron density is the expectation value of the number operator, given by $n_{\sigma}(i)=\langle \hat{n}_{\sigma}(i) \rangle$. The last term represents the chemical potential $\mu$, which fixes the average electron density of the system as $\rho=(1/N)\sum_{i,\sigma} n_{\sigma}(i)$, where $N$ is the total number of lattice sites. We also define the doping of the system as $\delta=(1-\rho)$. We conduct our study on a two-dimensional (2D) square lattice at low temperature $T\rightarrow0$, where the inverse temperature is denoted by $\beta=1/T$, with the Boltzmann constant set to unity.

We introduce an impurity potential at a specific lattice site $i_0$ by adding the following term to $\mathcal{H}$:
\begin{align}
\mathcal{H}_V= V \sum_\sigma c^\dagger_{\sigma}(i_0)c_{\sigma}(i_0)
\end{align}
here $V$ represents the impurity strength. We also consider a line impurity where a series of the same impurity is placed along a line $l$ along one axis
\begin{align}
\mathcal{H}_V= V \sum_{i\in l,\sigma} c^\dagger_{\sigma}(i)c_{\sigma}(i)
\end{align}
Our goal is to extract the retarded single-particle Green's function and thereby compute single-particle observables in the strongly correlated regime within the paramagnetic sector, i.e., for $U\gg t$.
%\begin{align}
%    \mathcal{G}_{\sigma \sigma^\prime}(i,j,\tau,\tau^\prime)=- \theta_H(\tau-\tau') \langle \{ c_{\sigma}(i, \tau ) , c_{\sigma'}^\dagger(j,\tau ')  \} \rangle_{\mathcal{H}} 
%\end{align}
%here $\theta_H(\tau-\tau')$ is the Heaviside function and $\langle ... \rangle_{\mathcal{H}}$ is the expectation value taken with the Hamiltonian $\mathcal{H}$. For simplicity, we will remove the subscript $\mathcal{H}$ in the subsequent discussions. 

\subsection{Composite operator method}
\label{sec:Method}
In this work, we employed an equation-of-motion technique to study the approximate ground state of Eq.~(\ref{eq:eqHubbard}) in the strongly correlated regime. We implement the method fully in real-space to study disorder, in the presence of which the translation symmetry is absent.

The composite operator method (COM)~\cite{avella2011composite} corresponds to the `first-order' equation of motion for Hubbard operators~\cite{hubbard1963electron,hubbard1964electron,tong2015equation,fan2018projective,jia2025thermal}. A detailed and systematic review and limitations of the COM has been recently provided by some of us~\cite{haurie2024bands,banerjee2024CDW}. Here, we provide a brief overview of the key assumptions of the method. Some additional details are presented in Appendix.~(\ref{App:COMDetails}).

The electronic annihilation operator $c_\sigma(i)$ can always be represented as the sum of two operators -- holons $\xi_\sigma(i)$ and doublons $\eta_\sigma(i)$
\begin{align}
c_\sigma(i)&=\xi_\sigma(i)+\eta_\sigma(i)
\label{Eq:cxieta}\\
\xi_\sigma(i)&=c_{\sigma}(i)\big(1-n_{\overline{\sigma}}(i)\big)
\label{Eq:xi}\\
\eta_\sigma(i)&=c_{\sigma}(i)n_{\overline{\sigma}}(i)
\label{Eq:eta}
\end{align}
Instead of studying the equation of motion of $c_{\sigma}(i)$, the idea is to study the equation of motion of $\xi_\sigma(i)$ and $\eta_\sigma(i)$. For the Hubbard model defined in Eq.~(\ref{eq:eqHubbard}), the equation of motions does not close for the hopping terms and has to be truncated to obtain a closed set of self-consistent equations. In the following, we assume that $\xi_\sigma(i)$ and $\eta_\sigma(i)$ are weakly interacting quasiparticle in the hole-doped Hubbard model. The truncation is followed by a projection of the new operators arising from the equation of motion procedure onto the $\xi, \eta$ basis. By making this approximation, we neglect non-local corrections. 
%In fact, if holons and doublons are the elementary charge excitations of a weakly doped Mott insulator, the bare holons and doublons of the atomic limit should be renormalized by these non-local effects.%If $\xi$ and $\eta$ are entirely incoherent in the Hubbard model, our approximation will not accurately capture the low-energy excitations of its ground state. \\ 

\subsection{Fourier transform and density of states\label{Sec:Fourier}}
The study of Friedel oscillation requires the computation of the real space dependence of the density of states $\rho(\omega,x)$ which is related to the composite Green's function via
\begin{align}
\rho(\omega,x)&=\dfrac{1}{L}\sum\limits_k\dfrac{-1}{\pi}\Im\left[\mathcal{G}_e(\omega,x,k)\right].&
%&=\dfrac{1}{L}\sum\limits_k\sum\limits_{\alpha\beta}\dfrac{-1}{\pi}\Im\left[\mathds{G}_{\alpha\beta}(\omega,x,k)\right]&
\end{align}

The local variation of the density of states $\delta\rho(\omega,x)$ is obtained by removing the site-averaged value,
\begin{align}
&\delta\rho(\omega,x)=\rho(\omega,x)-\dfrac{1}{L}\sum_{x_i}\rho(\omega,x_i).&
\end{align}
To analyze the oscillation frequency of the local density of states (LDOS), we compute the Fourier transform of $\delta\rho(\omega,x)$. The transform is performed using lattice sites that are at least $l$ lattice spacings away from the impurity, with $l_x=5$ chosen in practice to minimize local effects near the impurity. 
%\begin{align}
%\delta\rho(\omega,q)=\sum\limits_{|x_i|>l}e^{iqx_i}\delta\rho(\omega,x_i)
%\end{align}

\section{Results\label{sec:Results}}
We present self-consistent results in the presence of both a point-like ($0d$) impurity and a line ($1d$) impurity. For the point-like impurity, the translation symmetry is completely broken and the  calculations are performed on a $40\times 40$ square lattice. In the case of the line impurity, assuming translational symmetry along the impurity direction, Friedel oscillations can be computed for square lattices up to $512\times 256$ in size. Their dependence on $U$ and $n$ has been studies for $128\times 128$ lattice size. This size has also been used  for computing the Mott phase diagram. Our investigation spans doping levels from $n=0.86$ to $n=0.94$, at a fixed temperature of $T=0.01t$. 

\subsection{Friedel oscillations for low-impurity strength\label{Sec:LowImp}}
\begin{figure}[h!]
\includegraphics[width=0.49\textwidth]{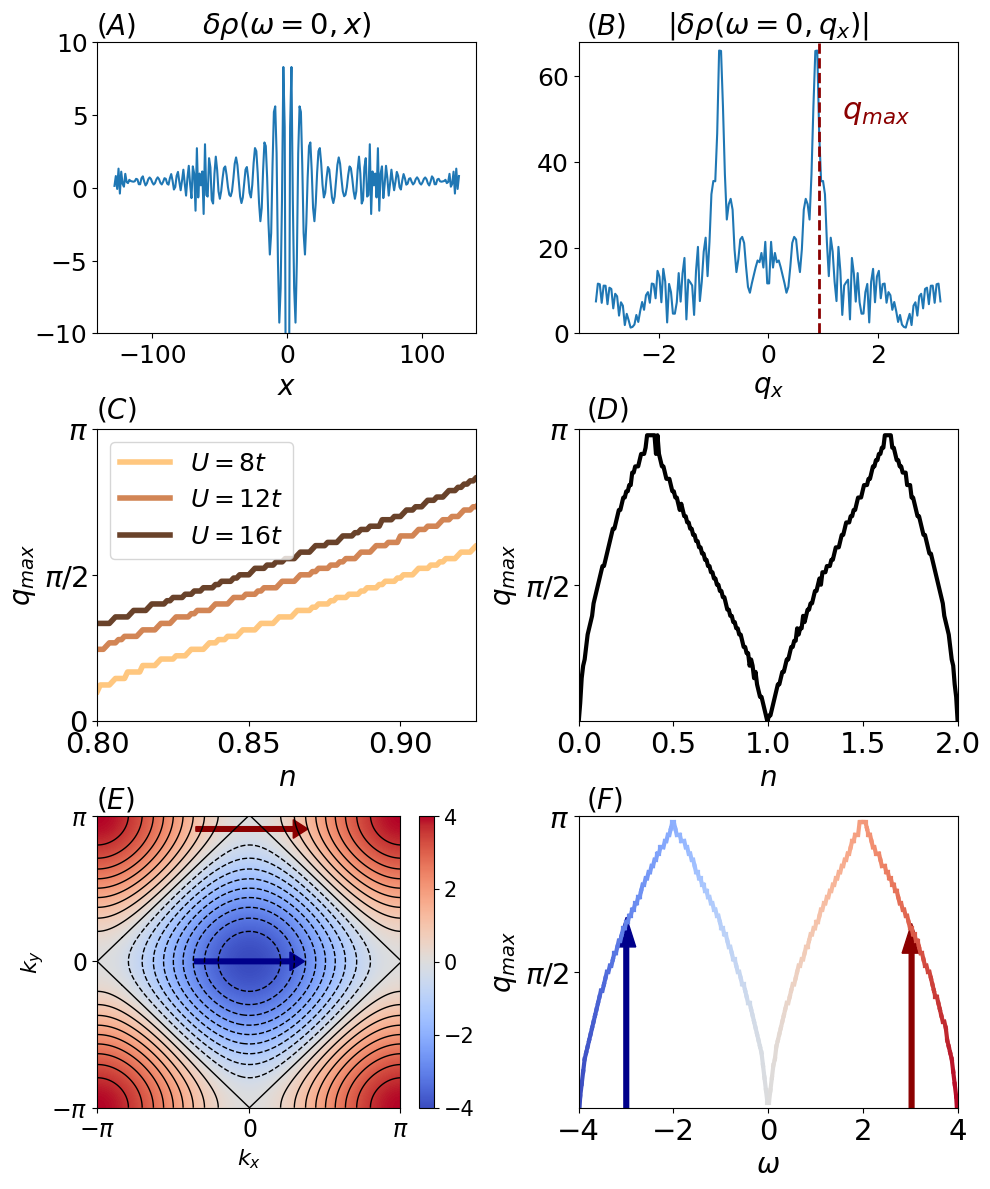}
\caption[0.5\textwidth]{(A) Real-space distribution of the density of states at the Fermi level in the presence of a repulsive line impurity located at $x = 128$, for parameters $U = 8t$, $\delta = 0.16$, and impurity strength $V_{imp} = 1t$ (B) Absolute value of the Fourier transform, $|\delta\rho(\omega=0, q_x)|$, exhibiting a distinct peak at $q_{max}$, marked by the red dotted line, indicating the characteristic frequency of the Friedel oscillations. (C) Dependence of the Friedel-oscillation frequency $q_{max}$ at the Fermi level ($\omega=0$) on the electron density $n$ for various interaction strengths: $U = 8t$, $U = 12t$, and $U = 16t$. (D) Friedel-oscillation frequency as a function of the electronic density $n$ for a non-interacting band insulator at the Fermi level. (E) Equal energy contours for a non-interacting band insulator at half-filling. (F) Frequency of the Friedel oscillations for a non-interacting band insulator as a function of energy, with colors corresponding to the equal-energy contours shown in (E). The Friedel oscillation frequencies at positive and negative energies are marked by blue and red arrows respectively. The corresponding Fermi surface scattering vectors are illustrated in panel (E).}
\label{fig:fig1}
\end{figure}

In Fig.~\ref{fig:fig1}(A), we show the local density of states $\delta\rho(\omega=0, x)$ at the Fermi level in the presence of a repulsive impurity line located at $x = 0$. Pronounced oscillations are observed across a broad range of Hubbard interaction strengths $U$ and electronic densities $n$, for both repulsive and attractive impurity potentials. To obtain the QPI pattern we perform a Fourier transform of the spatial spectra, as detailed in Sec.~\ref{Sec:Fourier}. We identify a well-defined peak, corresponding to the Friedel oscillation wavevector, $q_{\mathrm{max}}$ [Fig.~\ref{fig:fig1}(B)]. Remarkably, this wavevector remains sharply defined despite the strong electron correlations in the system, for which the hopping amplitudes and the local chemical potential are renormalized through self-consistent feedback. Furthermore, the amplitude of $q_{\mathrm{max}}$ is largely insensitive to the impurity strength.

Figure~\ref{fig:fig1}(C) illustrates the dependence of $q_{\mathrm{max}}$ on both the electronic density $n$ and the Hubbard interaction $U$. At fixed doping levels, the oscillation frequency $q_{\mathrm{max}}$ increases monotonically with $U$. Likewise, at fixed $U$, $q_{\mathrm{max}}$ grows with increasing electronic density within the range $n \in [0.80, 0.94]$. We compare this with the corresponding result in the absence of interactions, limit that can be solved exactly, and for which we have access to a much larger range of dopings (see Fig~\ref{fig:fig1}(D), for which we plot $q_{\mathrm{max}}$ as a function of the electronic density at the Fermi level ($\omega=0$)). We note that the results are quite different in the doping range accessible for the Mott insulator.

In order to understand the origin of this difference, in what follows we focus on the variation of $q_{\mathrm{max}}$ with energy for a fixed doping level. Thus, in Figure~\ref{fig:fig1}(F), we plot the frequency of the Friedel oscillations at half filling as a function of energy. In this non-interacting case, it is clear that varying the energy or doping yields the same result, as illustrated in Fig.~\ref{fig:fig1}(D/F). The Friedel oscillation wavevectors for a non-interacting system at a given energy are determined by the distance between certain regions of the equal-energy contours, for which the quasiparticle scattering is strongest. It has been argued that these regions correspond to the regions that have the largest curvature in momentum space~\cite{hoffman2002imaging,wang2003quasiparticle}. In Figs.~\ref{fig:fig1}(E) and (F) we plot these equal-energy contours(E), alongside $q_{\mathrm{max}}$ (F), using the same energy color scales. For two specific energy values we indicate by arrows the correspondence between the scattering regions in momentum-space, and the resulting FO wavevector.

%In Figure~\ref{fig:fig1}(D) we plot $q_{\mathrm{max}}$ as a function of the electronic density $n$ for a non-interacting band insulator at the Fermi level ($\omega=0$)
%To compare this result with the one corresponding to the non-interacting limit, in Figure~\ref{fig:fig1}(C) we plot the dependence of $q_{\mathrm{max}}$ at the Fermi level ($\omega=0$) as a function of doping. Since this limit can be solved exactly, we have now access to a much larger range of dopings. The results for the interacting and non-interacting case appear very different. In order to understand the origin of this difference, in what follows we focus on the variation of $q_{\mathrm{max}}$ with energy for a fixed doping level. For non-interacting systems, fixing the doping and varying the energy is exactly equivalent to fixing the energy and varying the doping. In Figure~\ref{fig:fig1}(D) we plot $q_{\mathrm{max}}$ as a function of the electronic density $n$ for a non-interacting band insulator at the Fermi level ($\omega=0$), while in Figure~\ref{fig:fig1}(F) we plot the frequency of the Friedel oscillations at half filling as a function of energy. We note that the two are indeed identical.

\begin{figure}[h!]
\includegraphics[width=0.49\textwidth]{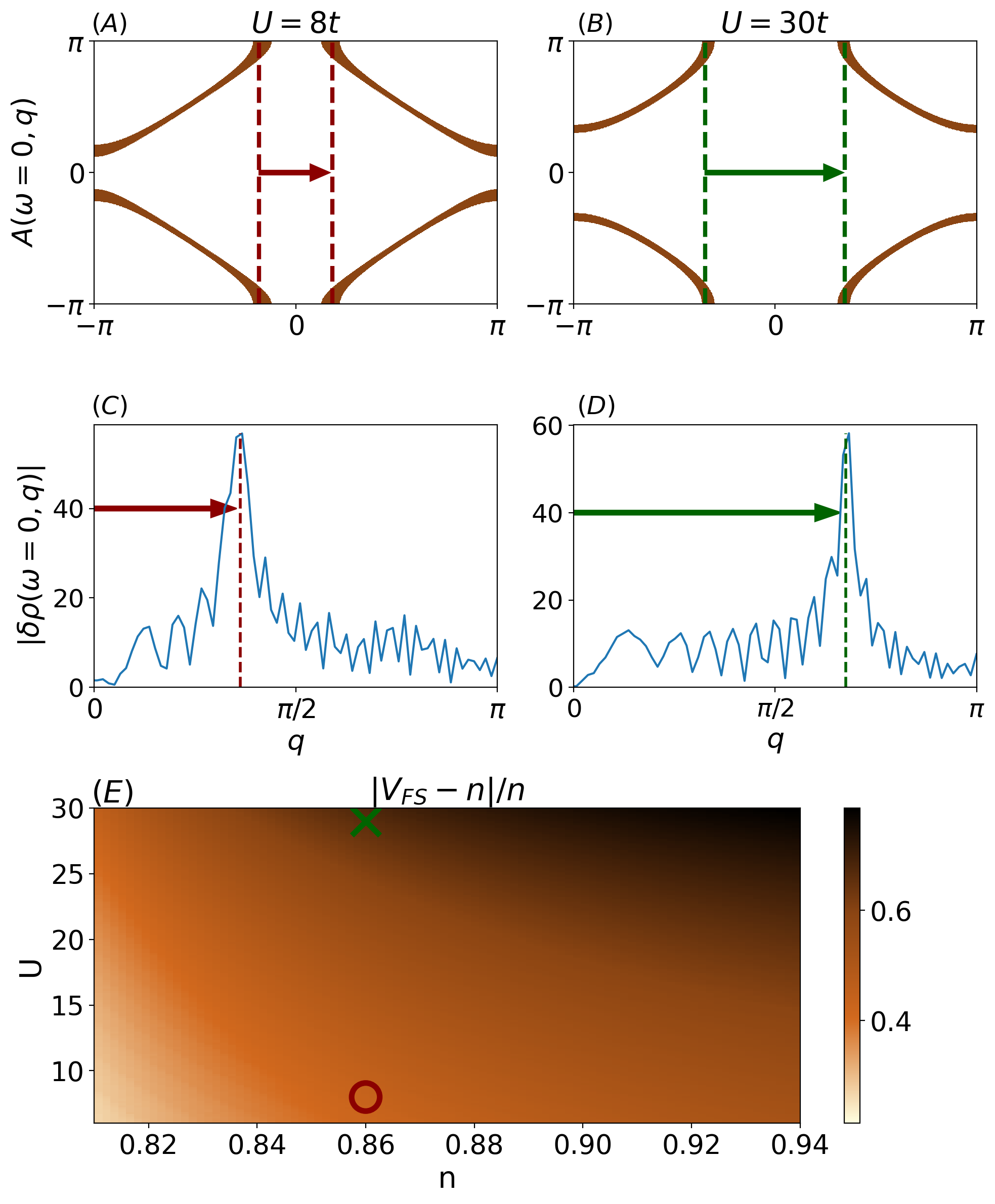}
\caption[0.5\textwidth]{(A/B) Fermi surface of a clean doped Mott insulator at doping $\delta = 0.1$ for (A) $U = 8t$ and (B) $U = 30t$. The red and green dotted lines indicate the minimal scattering wavevector associated with a line impurity. (C/D) The corresponding Fourier transform of the local density of states at the Fermi level for $\delta = 0.1$, in the presence of a repulsive line impurity with strength $V_{imp} = 1t$, for (C) $U = 8t$ and (D) $U = 30t$. The red and green arrows indicate the separation between the Fermi surface regions contributing to scattering, as shown in (A) and (B). (E) The ratio of the Fermi surface volume calculated using the COM approach to the Fermi surface volume predicted by Luttinger's theorem. This ratio highlights the extent of the violation of Luttinger's theorem as a function of the Coulomb repulsion $U$ and electronic density $n$.}
\label{fig:fig2}
\end{figure}

We expect a similar behavior for the doped Mott insulator, i.e a $q_{\mathrm{max}}$ that connects regions of the Fermi surface with the largest curvature. In Figs.~\ref{fig:fig2}(A) and \ref{fig:fig2}(B) we show the electronic Fermi surfaces at two different interaction strengths $U$ for a fixed filling $n = 0.9$, while the corresponding momentum-resolved density oscillations $\delta\rho(\omega, q)$ are shown in Figs.~\ref{fig:fig2}(C) and \ref{fig:fig2}(D). The Friedel oscillation $q_{\mathrm{max}}$ corresponds indeed to the wavevector connecting the same regions of the Fermi surface as in the non-interacting case, as indicated by the arrows in Figs.~\ref{fig:fig2}(A, B). 

We can see that, despite the non-linear, self-consistent nature of the composite operator formalism, weakly doped Mott insulators with weak impurities behave effectively as a system of weakly interacting holons. In this regime, Friedel oscillations are controlled by the shape and volume of the `holon Fermi surface'. For a conventional Fermi liquid, the Fermi surface volume $V_{\mathrm{FS}}$ is determined by Luttinger theorem~\cite{luttinger1960fermi}, which states:
\begin{align}
\frac{V_{\mathrm{FS}}}{(2\pi)^d} = n \mod 2\pi
\end{align}
where $n = N/L^d$ is the electronic density in $d$ dimensions. However, in Figs.~\ref{fig:fig2}(A,B), we observe that the reconstructed Fermi surface encloses a volume larger than that predicted by Luttinger's theorem. This violation is quantified in Fig.~\ref{fig:fig2}(C), which shows the deviation as a function of $n$ and $U$. Compared to a non-interacting system, the resulting discrepancy in Friedel oscillation wavevectors reflects a fundamental breakdown of Luttinger's theorem near the Mott insulator phase.

While violations of Luttinger's theorem within the Hubbard operator framework are routinely reported~\cite{osborne2021broken,haurie2024bands}, the underlying origin of this anomaly remains a topic of ongoing debate.  In the hole-doped regime, the composite operator method treats holes as the charge carriers, and such holes arising from a half-filled Mott insulating state do not obey fermionic algebra. As we demonstrate in Appendix~\ref{App:LuttingerViolation}, this non-fermionic algebra inherently leads to a breakdown of Luttinger's theorem.

Near the Mott insulating phase, the electronic degrees of freedom decompose into holon and doublon excitations, leading to a reduced spectral weight for holons. In a paramagnetic state, the expectation value
\begin{align}
\langle \{ \xi_{\sigma}(i), \xi^\dagger_{\sigma}(i) \} \rangle = 1 - \frac{n(i)}{2}
\end{align}
determines the weight of the lower Hubbard band. The deviation of this quantity from unity signals a departure from Fermi-liquid behavior, with the maximal deviation occurring at half-filling. 

In the large-$U$ limit, only the holon band is occupied in a hole-doped Mott insulator. These holons can be treated as weakly correlated quasiparticles with renormalized effective mass and reduced spectral weight. Consequently, the sum over occupied holon states equals the hole density, suggesting a modified Luttinger's theorem for holons. However, since the spectral weight is less than unity, the holon Fermi surface expands to accommodate all $N$ electrons, resulting in a violation of the conventional Luttinger's theorem.

The dependence of the Luttinger violation, shown in Fig.~\ref{fig:fig2} (E) as a function of $U$ and $n$, can now be qualitatively understood. As the doping increases, the weight of holons approaches that of non-interacting electrons, reducing the Luttinger violation. In contrast, increasing $U$ enhances the violation. At intermediate interaction strength, residual holon-doublon hybridization softens the anomaly. However, as $U$ becomes large and the Hubbard bands separate, this hybridization is suppressed, leading to a pronounced breakdown of Luttinger's theorem~\cite{harris1967single,phillips2010colloquium}.

\subsection{Phase separation at large-impurity strength\label{Sec:LargeImp}.}
In the previous section we have shown that Friedel oscillations can be observed in a doped Mott insulator for weak impurities, akin to the non-interacting system. However, the dominant wavevector of the Friedel oscillation is modified due to the violation of Luttinger's theorem in the proximity of the Mott insulator regime. Such violation of the Luttinger's theorem originates from the non-fermionic nature of the charge excitations of the parent Mott insulator state. 

In what follows we focus on larger impurity strengths, and we find that, when the impurity strength exceeds a critical value, the system transitions into a phase-separated state  This arises from the instability of the uniform doped Mott insulator, when treated within the Hubbard operator formalism, as shown in a previous study~\cite{pangburn2024spontaneous}. Two distinct spatially separated regions emerge in the phase-separated regime: a Mott region around the impurity for which the electronic density remains fixed and equal to unity, and a hole-rich region.

\subsubsection{Friedel oscillations in the presence of a strong impurity}
\begin{figure}[h!]
\includegraphics[width=0.49\textwidth]{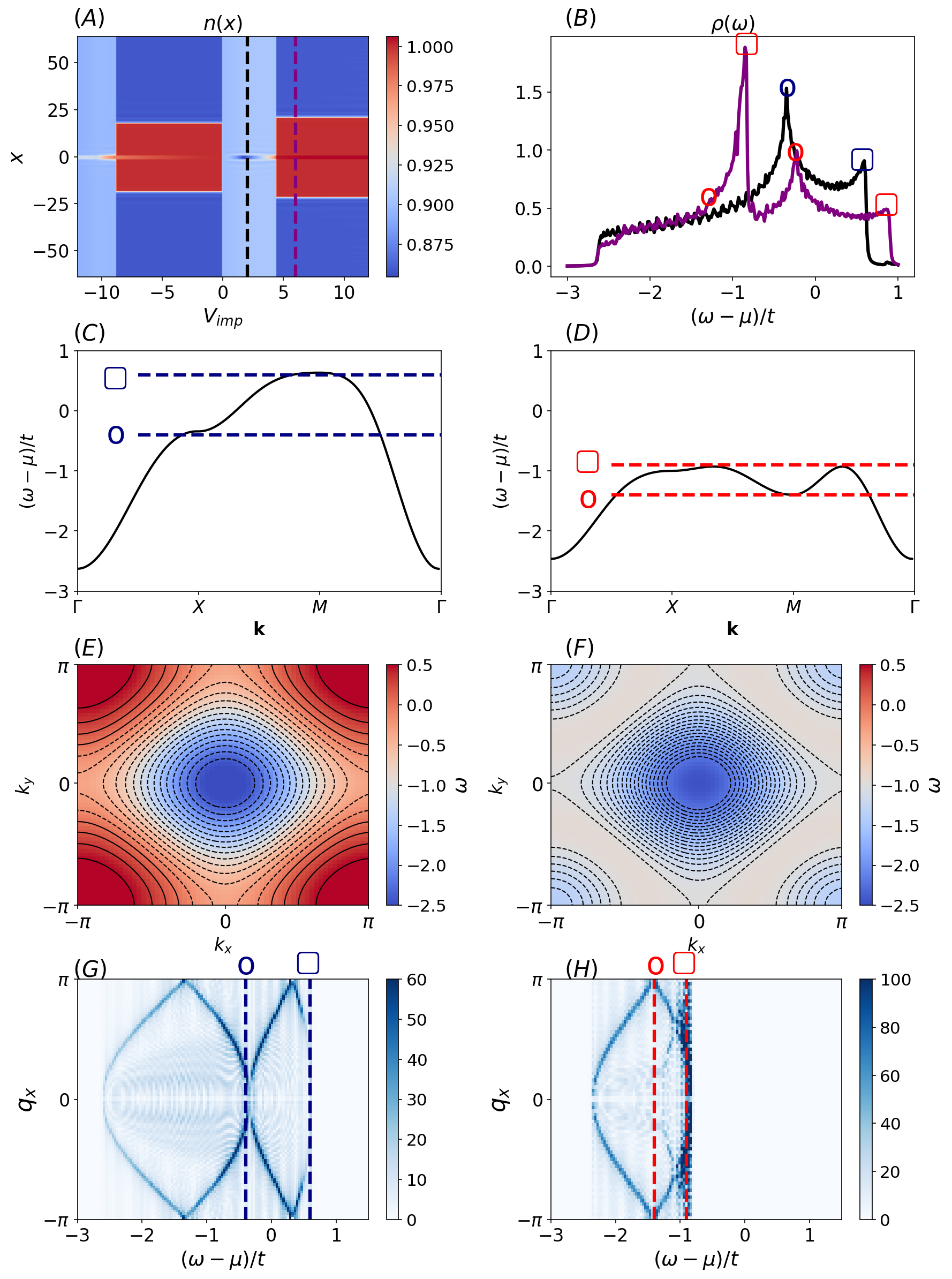}
\caption[0.5\textwidth]{(A) The electronic density as a function of position and impurity strength for $U=12t$ and $n=0.90$. (B) The spatially-averaged density of states for impurity strengths below (black) and above (purple) the threshold for the Mott-region formation, corresponding to the values indicated by the black and purple lines in panel (A).  The peaks in the density of states (DOS) corresponding to the non-Mott contribution are marked by a blue square and a blue circle, and correspond respectively to the top of the band and the Van-Hove singularity, as indicated in the corresponding band structure in panel (C). The Mott region contribution to the DOS consists in two peaks corresponding to the top of the Mott band (red square), and to the Van-Hove singularity (red circle), as indicated also in the band dispersion (D). (E/F) Equal-energy contours for a clean system with $U=12t$, for $n=0.90$ (E) and $n=1.0$ (the Mott phase) (F). (G/H) The Fourier transform of the LDOS, $\delta\rho(\omega,q)$ for values of the impurity strengths corresponding to the black and purple lines, i.e. with and without formation of the Mott regions. The special points in the band structure are indicated by the dashed lines.}
\label{fig:fig3} 
\end{figure}

In Fig.~\ref{fig:fig3} (A) we plot the electronic density as a function of position and impurity strength for $U=12t$ and $n=0.90$. Here $x=0$ represents the position of the impurity line. Mott regions ($n=1$, denoted in red) form in the vicinity of the impurity for certain values of the impurity strength, for both attractive and repulsive impurities. Note that, in the repulsive case, the Mott region forms for impurity values larger than a critical threshold, while in the attractive one, the phase-separated phase arises as soon as we turn on the impurity, and disappears at very large values of the impurity potential. 
%However, for a given density and on-site repulsion $U$, the ratio of the size of the Mott region to the size of the system {\bf check} remains constant with the impurity strength. 

In Fig.~\ref{fig:fig3} (B), we compare the spatially-averaged density of states for the non-phase-separated regime (black) and the phase-separated one (purple), corresponding to the impurity strengths indicated by the black and purple lines in Fig.~\ref{fig:fig3} (A). For the non-separated phase, $\rho(\omega)$ is similar to that previously obtained for the lower Hubbard band of a uniform system~\cite{haurie2024bands} at filling $n=0.90$, as expected. The corresponding band structure is depicted in Fig.~\ref{fig:fig3} (C). Note that the lower Hubbard band crosses the Fermi energy. Also, note formation of two peaks in the DOS: the first one, at $E \approx 0.7$ corresponds to the top of the band and is indicated by a blue square,
%, as depicted in dark red in Fig.~\ref{fig:fig3} (C) where we plot the equal energy contours of the lower Hubbard band. 
while the second one at $E \approx -0.7$, indicated by a blue circle, corresponds to a Van-Hove singularity, which in Fig.~\ref{fig:fig3} (E) marks the transition between equal-energy contours centered around the center and the corners of the BZ.

For the impurity value corresponding to a phase-separated system (corresponding to the purple line), the average density of states takes into account both the Mott and non-Mott regions. We expect the contribution of the density of states of the non-Mott region to be similar to the one for the uniform system, and give rise to features similar to those depicted in the black-line plot. Indeed we note the formation of two peaks (denoted by a red square and a red circle) at similar energies to those corresponding to the uniform system (blue square and blue circle). The peak energies differ slightly between the phase-separated and uniform regimes, as the non-Mott region occurs at a different doping level than in the uniform case.

The contribution of the Mott region to the average DOS can be inferred by plotting the Mott band structure for a uniform system (Fig.~\ref{fig:fig3} (D)). Note that the lower Hubbard band in this regime has a smaller bandwidth and is centered at negative energies farther away from the Fermi level of the phase separated system. This confirms the gapped nature of local excitations in this regime and shows that inserting an electron into the Mott region requires an energy $\sim U$, due to double occupancy. Moreover, the bands below the Fermi level become significantly flatter, reflecting a suppressed kinetic energy in the Mott state. The corresponding equal-energy contours are depicted in Fig.~\ref{fig:fig3} (F). The main feature of this band structure is thus the quasi-flat band at $\omega\approx-1$ that corresponds to the top of the Mott band. This yields a strong peak in the DOS at the corresponding energy, which is indicated by the red square in Fig.~\ref{fig:fig3} (B). This peak represents the energy required to extract electrons from the Mott regions~\cite{banerjee2024CDW,pangburn2024spontaneous}. 

In Fig.~\ref{fig:fig3} (F) note  the formation of two types of equal energy contours, some centered around the corners, and some around the center of the BZ. The pockets around the corners disappear for energies around $\omega\approx-1.2$. Below this energy we only have contours centered around the center of the BZ. This energy gives also rise to a small peak in the DOS, denoted by a red circle, however, given the closeness of the two energy values it is hard to distinguish it from the $\omega\approx-1$ peak. In the band structure this corresponds to a band inflexion (Van-Hove singularity) marked also by a red circle.

%hen Mott regions form,reduces near $\omega=0$, whereas the quasiparticle peak at the negative energy strengthens. Such peak at negative energies represents the energy required to extract electrons from the Mott regions~\cite{banerjee2024CDW,pangburn2024spontaneous}.

While we have till now focused on the average DOS, in Fig.~\ref{fig:fig3} (G) we show the Fourier transform of the local density of states $|\delta\rho(\omega,q_x)|$. The special points in the band structure are marked by the vertical dashed lines.
For an impurity strength corresponding to the non-separated phase, $|\delta\rho(\omega,q_x)|$ shows a similar behavior to its counterpart for a non-interacting systems depicted in Fig.~\ref{fig:fig1} (F), with deviations arising from the different band structure for the lower Hubbard band, and the violation of the Luttinger theorem. 

In Fig.~\ref{fig:fig3} (H) we plot $|\delta\rho(\omega,q_x)|$ corresponding to the phase-separated regime. We perform a selective Fourier transform such that we take into account only the Mott region (depicted in red in Fig.~\ref{fig:fig3} (A)). Note that the Friedel oscillations in this region are generated both by the impurity scattering and by the existence of the sharp boundaries of the Mott region. As expected, they reflect the physics of the Mott band described in Figs.~\ref{fig:fig3} (D) and (F). Thus the FO extend in energy in a reduced interval centered at negative energies. We note a very strong feature at $\omega \approx -1$, as well as dispersing features corresponding to the energies in the rest of the band. Fig.~\ref{fig:fig3} (H) shows also some faint additional oscillations in the energy interval $-1.2 \lessapprox \omega \lessapprox -1$ which are due to the additional Fermi pockets shown in Fig.~\ref{fig:fig3} (D). The corresponding special points in the band structure are marked by the vertical red dashed lines.

%In Fig.\ref{fig:fig3}(E), we plot the equal-energy contours corresponding to the uniform phase at filling $n=0.90$, which match the average electronic density underlying the density of states oscillations shown in Fig.\ref{fig:fig3}(C). The band structure along high-symmetry lines at the same parameters is presented in Fig.~\ref{fig:fig3} (G), which shows the lower Hubbard bands cross the Fermi energy. On the other hand, for the phase-separated state, the equal-energy contours of the Mott region are presented in Fig.\ref{fig:fig3}(F) and the band structure in Fig.\ref{fig:fig3}(H). These confirm the gapped nature of local excitations and show that inserting an electron into the Mott region requires energy $\sim U$, due to double occupancy. Moreover, the bands below the Fermi level become significantly flatter, reflecting suppressed kinetic energy in the Mott state.

\begin{figure}[h!]
\includegraphics[width=0.49\textwidth]{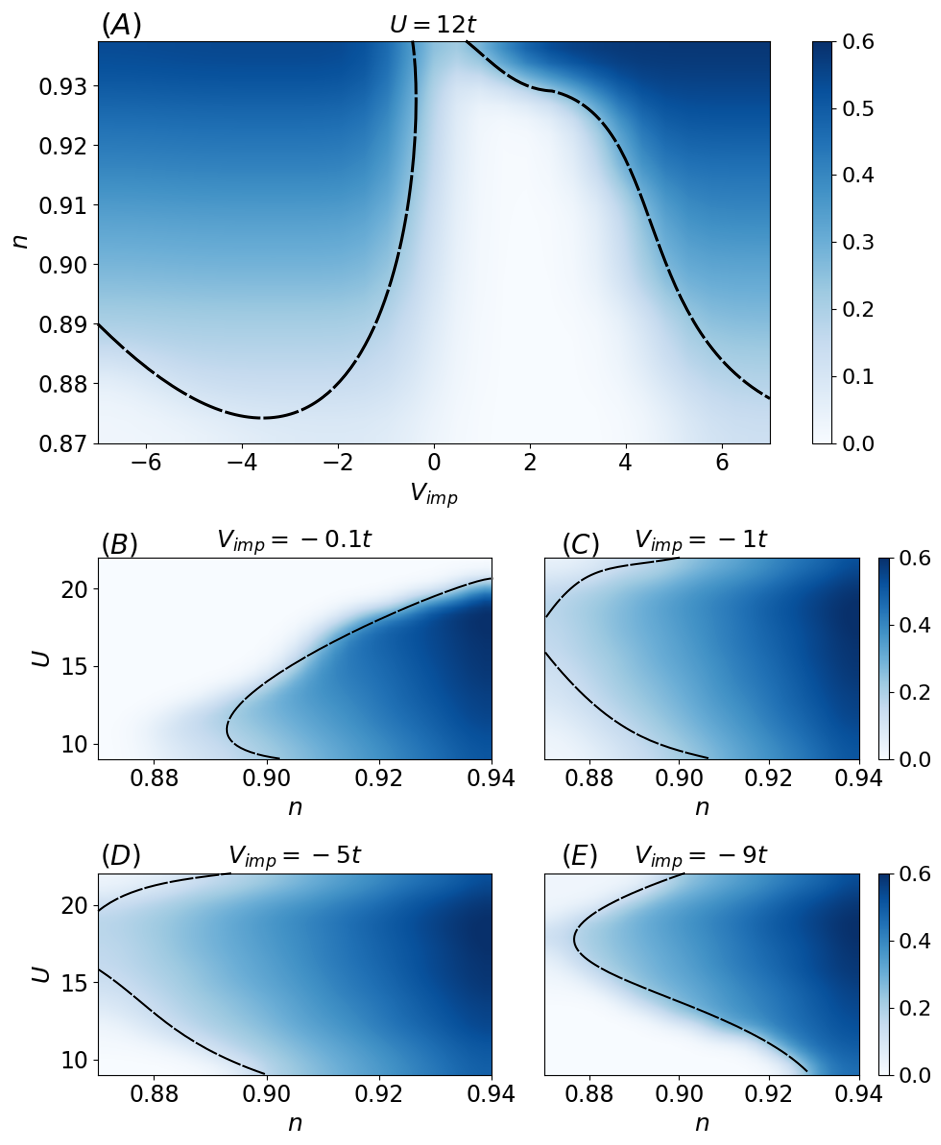}
\caption[0.5\textwidth]{(A)Fraction of space occupied by the Mott region as a function of the electronic density and impurity potential $V_{imp}$ for $U=12t$. (B-E) Fraction of space occupied by the Mott region  as a function of the electron density $n$ and the Hubbard interaction $U$, for varying strengths of an attractive line-impurity with (B) $V_{imp}=-0.1t$, (C) $V_{imp}=-1t$, (D) $V_{imp}=-5t$, and (E) $V_{imp}=-9t$.}
\label{fig:fig5}
\end{figure}

\subsubsection{Phase diagram}
Fig.~\ref{fig:fig5} (A), shows the evolution of the fraction of the half-filled Mott regions as a function of the electronic density and impurity potential $V_{imp}$ for $U=12t$. In Figs.~\ref{fig:fig5}(B)--(E), we present the corresponding phase diagrams in the $U$--$n$ plane for different impurity strengths. Near half-filling the Mott regions dominate large portions of the system, up to about $60\%$. Below some critical threshold for the doping, indicated by the dotted lines, and dependent on $U$ and $V_{imp}$, the system can no longer support phase-separated Mott regions.

Interestingly, the stability of these phase-separated regions exhibits a non-monotonic dependence on both $U$ and the impurity strength, as shown in Figs.~\ref{fig:fig5}(B--E). This behavior reflects a competition between the kinetic energy of mobile holes and the energy gain from forming tightly bound holon-doublon pairs in the Mott regions. The formation of such pairs tends to expel holes from the Mott regions, enhancing their insulating character. We discuss this in detail Sec.~(\ref{Sec:MicroOrigin}).

\subsubsection{Point Impurity}
\begin{figure}[h!]
\includegraphics[width=0.49\textwidth]{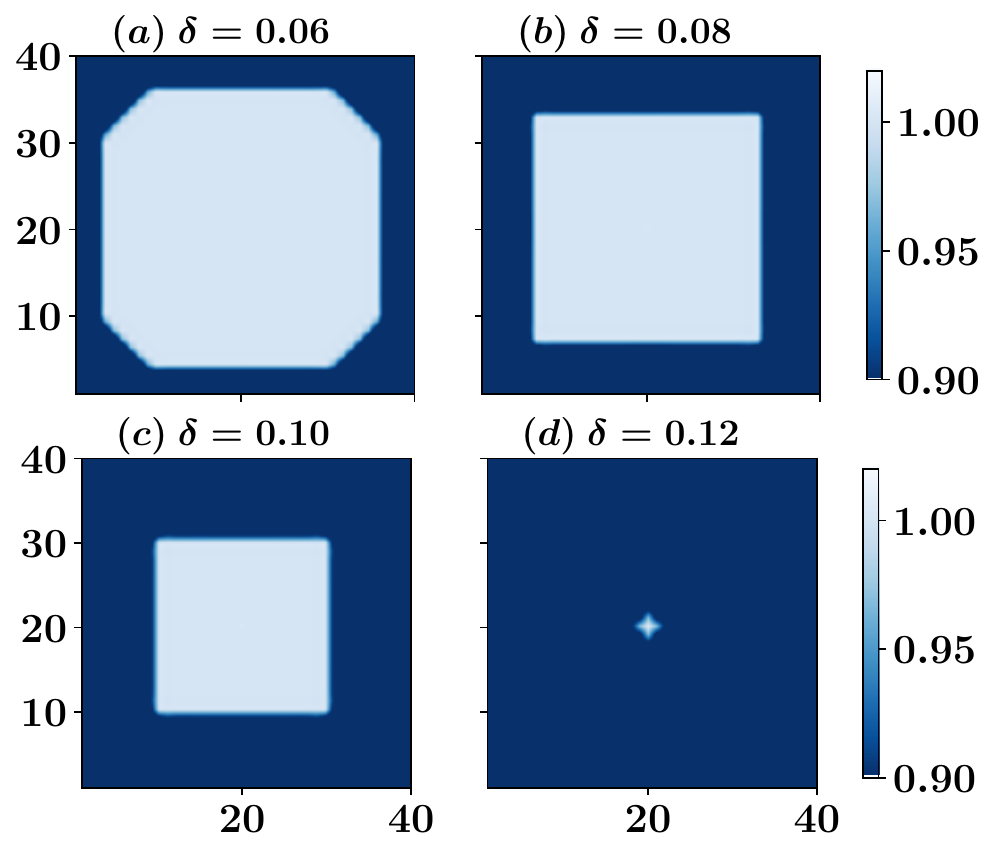}
\caption[0.5\textwidth]{Four representative spatial profiles of the self-consistent electron density with a single impurity at the center $r_0=(20,20)$: (a)  $\delta=0.075$ and $V_0=-0.2t$ (b) $\delta=0.1$ and $V_0=0.5t$ (c) $\delta=0.125$ and $V_0=-0.6$ (d) $\delta=0.140$ $V_0=-1.0t$ Note the Mott region created around the impurity. The size of this region reduces as doping increases.}
\label{fig:fig6}
\end{figure}

In what follows, instead of a system with a line impurity we focus on a a point impurity. Due to the breaking of translational symmetry in both spatial directions, we now need to solve the full self-consistent equations in a two dimensional system. In Fig.~\ref{fig:fig6} we plot the spatial profile of the electron density in the presence of a single attractive impurity located at the center of a $ 40\times 40$ system. At low doping, a large Mott region emerges in the vicinity of the impurity site. Same as for the line-impurity case, as the doping level increases, the extent of the Mott region diminishes significantly. Notably, for doping levels $\delta > 0.12$, the Mott region fails to form at interaction strength $U = 8t$. 

%This observation highlights the fact that the emergence of Mott regions is not solely a local effect of the impurity. Instead, it manifests as a global phase-separation instability inherent to the doped Hubbard model. Introducing a local impurity explicitly breaks translational symmetry, thereby acting as a perturbation that triggers the formation of a Mott region. 

\begin{figure}[h!]
\includegraphics[width=0.49\textwidth]{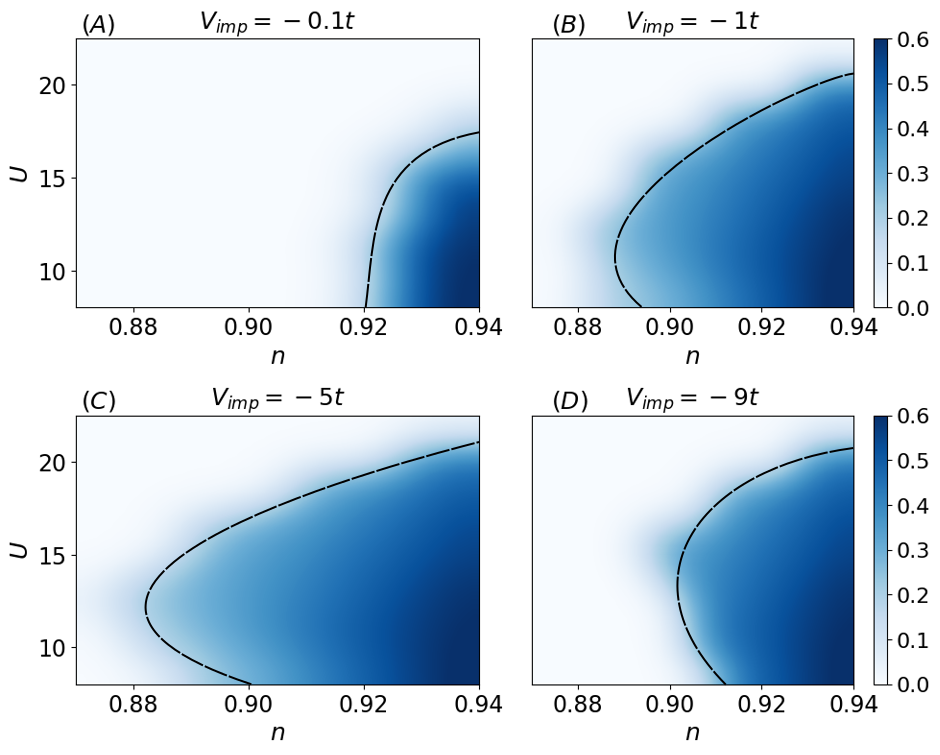}
\caption[0.5\textwidth]{Fraction of the Mott region as a function of the electron density $n$ and the Hubbard interaction $U$ for a point-like attractive impurity with different impurity values (A) $V_{imp}=-0.1t$, (B) $V_{imp}=-1t$, (C) $V_{imp}=-5t$ and (D) $V_{imp}=-9t$.}
\label{fig:fig9}
\end{figure}

In Figure~\ref{fig:fig9}  we present the phase diagram showing the fraction of the Mott region in the $U$-$n$ plane for a point impurity with different values of the impurity potential. While the overall trends qualitatively resemble those in Fig.~\ref{fig:fig5} obtained for a line impurity, the parameter space supporting the formation of a Mott region formation is significantly reduced in the point-impurity case, as it can be intuitively expected given the lower dimensionality of the impurity with respect to the uniform system.

Same as for the line-impurity case, the size of the Mott region decreases systematically with increasing doping and interaction strength $U$, and, at high doping, the phase-separated Mott state can no longer be stabilized. This suppression arises because the increasing hole concentration diminishes the energetic advantage of forming a Mott region. These results are consistent with recent observations in Ref.~\cite{banerjee2024CDW}, where a charge-density-wave phase emerging in both the Hubbard operator formalism and the related $t$-$J$ model~\cite{BanerjeetJ} vanishes beyond a critical doping threshold.

The dependence of Mott region formation on $U$ is non-monotonic. As $U$ increases from $8t$ to $22t$, the Mott regions initially grow, reflecting an enhanced localization. However, beyond a certain point, a further increase in $U$ suppresses the phase-separated state. This nontrivial behavior arises from the competing influences of holon-doublon hybridization and holon mobility, a mechanism we explore in detail in the subsequent section.

The size of the Mott region also depends on the impurity strength, for both attractive and repulsive cases, as shown in Fig.~\ref{fig:fig5} and \ref{fig:fig9}. For attractive impurities, when the impurity potential becomes strong enough that the local electronic density approaches one, a Mott instability arises. This leads to the formation of a Mott region around the impurity, as double occupancy is suppressed due to the large on-site Coulomb repulsion $U$. However, when the impurity strength becomes comparable to or exceeds the interaction energy ($V_{imp}\sim U$), double occupancy at the impurity site becomes possible, which destabilizes the surrounding Mott region.
%At weak impurity strength, increasing the impurity potential allows the phase-separated state to persist at higher doping levels. 
%To get an intuitive understanding of this phenomenon we note that a Mott instability arises when the electronic density at a given site reaches unity. In the presence of an attractive impurity, the impurity site rapidly accumulates electrons and reaches half-filling, thereby inducing a Mott instability even at low impurity strengths. 
Conversely, a repulsive impurity depletes the electronic density at the impurity site. However, due to Friedel oscillations, the density in the neighboring regions can reach unity, eventually triggering a Mott instability, but at higher impurity strengths compared to the attractive case.
%In the attractive case, the impurity site must reach $n=1$ to nucleate a Mott region, whereas in the repulsive case, the surrounding sites must reach half-filling. A stronger impurity helps achieve these density conditions, facilitating the formation of the Mott region. 
Once established, the size of the Mott region becomes largely independent of the impurity itself, indicating that the impurity acts primarily as a trigger for an underlying phase-separated instability in the doped Hubbard model.

%However, when the impurity potential becomes comparable to or exceeds the Coulomb repulsion ($V_{imp} \sim U$), double occupancy becomes energetically favorable at attractive impurity sites, simultaneously, nearby sites lose electrons due to Friedel oscillations, leading to local densities below half-filling, consequently, the size and stability of the Mott region diminishes during this regime.

%In summary, forming Mott regions requires a sufficiently strong impurity potential for both point and line impurities. The Mott region expands with increasing electronic density, but the phase-separated state becomes unstable beyond a critical hole doping. For fixed impurity strength, the stability of the Mott region exhibits a non-monotonic dependence on $U$, and excessive impurity potentials reduce Mott stability due to the onset of double occupancies. In the next section, we analyze the microscopic origin of the phase-separated Mott state in the doped Hubbard model.

\subsubsection{Origin of the Mott-separated phase}
\label{Sec:MicroOrigin}
\begin{figure}[h!]
\includegraphics[width=0.43\textwidth]{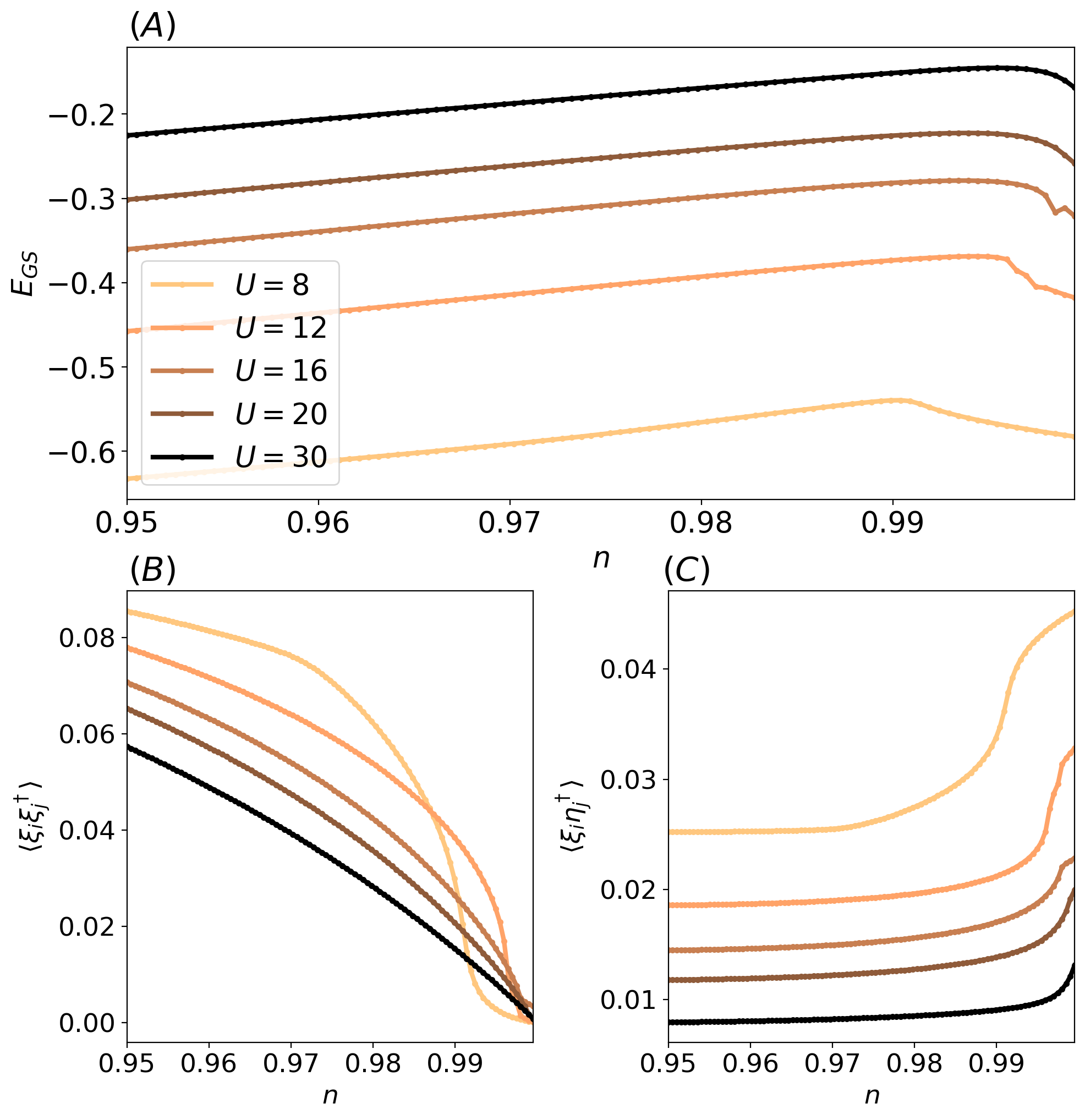}
\caption[0.5\textwidth]{For different repulsion strengths $U$ we plot (A) The energy per site as a function of electron density $n$ (B) The hole mobility $\langle \xi(i)\xi^\dagger(j)\rangle$ as a function of $n$ (C) The holon-doublon hybridization $\langle \xi(i)\eta^\dagger(j)\rangle$ as a function of $n$.}
\label{fig:fig8}
\end{figure}

Figure~\ref{fig:fig8}(A) shows that the energy per site, $E_s(\delta)$, for a uniform solution of the one-band Hubbard model, exhibits a non-monotonic behavior as a function of the hole doping $\delta$. This non-monotonicity near half-filling suggests that phase-separated states are energetically favorable in this regime, consistent with earlier findings~\cite{pangburn2024spontaneous}. Moreover, the extent of this non-monotonicity decreases with increasing the Hubbard repulsion $U$, indicating that the energy gain from phase separation diminishes at larger $U$ and higher doping. As a result, the system transitions to a uniform state beyond a critical hole doping.

To understand the origin of this behavior, we analyze the key contributions to $E_s(\delta)$. Figure~\ref{fig:fig8}(B) displays the doping dependence of hole mobility, which increases with both doping and $U$. Near half-filling and at moderate values of $U$, electrons predominantly move through holon-doublon mixing processes. In this regime, the mobility of both holes and doublons is suppressed due to their tendency to form bound holon-doublon pairs, while the strength of holon-doublon hybridization increases. This scenario is reflected in the nearest neighbor holon-doublon hybridization shown in Fig.~\ref{fig:fig8} (C). %Notably, hybridization decreases with increasing $U$, due to the growing energy gap between holon and doublon bands. 
Interestingly, such mixing decreases as a function of $U$ due to the significant energy gap between the holon and doublon bands. This trend highlights the competition between holon mobility and holon-doublon pairing. Close to half-filling, isolated holon motion is strongly suppressed, while holon-doublon mixing becomes the dominant kinetic process. We further confirm that in the hole-doped regime, doublon mobility, quantified by $\langle \eta(i)\eta^\dagger(j) \rangle$, remains negligible and does not contribute significantly to the dynamics.

Based on these observations, we can conclude that near half-filling and for intermediate values of $U$, the system energetically favors the formation of fluctuating, tightly bound holon-doublon pairs. These pairs maintain the average density at half-filling and can lower the total energy by gaining kinetic energy through nearest-neighbor hopping. This mechanism naturally leads to the emergence of Mott regions. However, as doping increases, holons become increasingly delocalized, suppressing holon-doublon pairing and driving the system toward a uniform metallic phase. Our calculations provide evidence of a dynamical, charge-paramagnetic ground state near half-filling for $U \gg t$, in contrast to a static Mott insulator where only spin degrees of freedom remain active. This charge-fluctuating state is favored because it enables kinetic energy gain despite strong interactions, as discussed in Refs.~\cite{hirsch1989bond,hirsch1992superconductors}.

\section{Discussion \label{Sec:Discussions}}

\subsection{Friedel oscillations and non-fermionic quasiparticles}
The Friedel oscillation results presented in this work rely on two key aspects: the existence of well-defined low-energy quasiparticles that couple to the impurity, and the non-fermionic statistics obeyed by these quasiparticles. While the decomposition of electrons into holons and doublons is exact, when truncating the equation of motion, the algebra is not treated exactly. The main limitation of the truncation lies in its bias toward Mottness~\cite{phillips2006mottness} as it originates from the atomic limit. As a result, the system remains in a doped Mott insulating state across all values of $n$, even though a transition from a doped Mott insulator to a weakly correlated metal is expected at a finite critical doping~\cite{liebsch2009finite,vidhyadhiraja2009quantum,sordi2010finite} leading to a Fermi suface reconstruction without symmetry breaking~\cite{gazit2020fermi}. This transition should be accompanied by an evolution of the Fermi surface from small to large volume~\cite{sachdev2018topological}. A possible way to solve this issue while keeping a simple basis of holons and doublons is to assume that higher-order operators neglected here can be integrated out, renormalizing the bare holons and doublons. In particular, these higher-order contributions would renormalize the $I$-matrix, hence modify the weights of holons and doublons. As a result, the main conclusions regarding Friedel oscillations remain valid for a more general framework, although the oscillation frequency would shift because of the corresponding shift in the degree of Luttinger theorem violation.

\subsection{Phase-separation in the Hubbard model}
Local impurities probe the low-energy excitation. However, the formation of the Mott regions surrounding the impurity destroys low-energy excitations as the excitations become gapped. Consequently, the Mott insulating regions effectively shields the impurity. Such formation of Mott regions been proposed in strongly disordered $t-J$ model~\cite{Debmalya1}.

The Hubbard model describes the physics of strongly correlated systems for which the long-range Coulomb interactions are screened. However, if the ground state of the Hubbard model results in a phase-separated state, the charge heterogeneity makes the long-range Coulomb interaction relevant~\cite{shenoy2009long}. Intuitively, long-range Coulomb smears out large-scale charge inhomogeneities and can fragment the phase-separated state into smaller, spatially distributed puddles. The relationship of such a frustrated phase-separated state with the stripes~\cite{banerjee2024CDW} in the Hubbard model remains a challenge for the future.

\subsection{Outlook}
Among the various physical mechanisms proposed for phase separation, the spin-bag~\cite{schrieffer1988spin} mechanism has been extensively studied in the context of antiferromagnetism. This involves the local suppression of magnetic order due to the presence of holes, which induces an effective attractive interaction between holes. It has been explored using nonlocal operators~\cite{eder1994spin,eder1995anomalous,wrobel1998symmetry} such as $\xi_{\sigma}(i)S^+(j)$, which explicitly couple charge and spin degrees of freedom at different sites. These studies suggest that phase separation can emerge from the disruption of local magnetic order, providing a route for hole pairing and clustering. This phenomenon has also been proposed more generally~\cite{hirsch1989pairing} as a fundamental consequence of doping a correlated insulating state, wherein the destruction of local order facilitates the spatial segregation of charge.

In contrast, our work proposes an alternative mechanism for phase separation that does not rely on long-range magnetic order. Instead, we consider a paramagnetic background and demonstrate that phase separation can arise from forming tightly bound holon-doublon pairs within the Mott regions. This mechanism highlights the role of charge fluctuations and local correlations in driving a phase separation, independent of magnetic ordering. Nevertheless, verifying the robustness of this picture in frameworks that incorporate nonlocal interactions and quantum fluctuations more accurately remains an important direction for future research.
 
 \subsection{Conclusions}
We investigated the formation of Friedel oscillations in the presence of a single localized line and point impurity in a doped Mott insulator described in the composite operator framework. We found that these oscillations resemble those observed in non-interacting systems, however their wavevector reflects the violation of Luttinger's theorem. The similarity with the non-interacting physics stems from the fact that the physics of this system can be described by an effective non-interactive model for the holon and doublon excitations. Moreover, at larger impurity strengths, the system may transitions to a phase-separated state with a Mott-insulating (half-filled) region forming in the vicinity of the impurity, and hole-rich regions farther away from it. We have studied the phase transitions associated with this phase separation and we identified the effect of each of the underlying parameters (doping, interaction strength, and impurity value) to the formation of the Mott phase.

\section{Acknowledgments}
The authors thank V. B. Shenoy, A. Ghosal and C. Mahato for useful discussions.  A.B. and C.P. acknowledges funding from CEPIFRA (Grant No. 6704-3). The calculations were performed on the IPhT Kanta cluster. 

\appendix
\section{Details of the composite operator formalism\label{App:COMDetails}}
Since the COM formalism has already been reviewed in previous works~\cite{haurie2024bands,banerjee2024CDW,pangburn2024spontaneous}, we present the underlying equations for completeness, without however detailing their derivation.

The composite operator method is an equation of motion method aiming at computing the single-particle electronic Green's function $\mathcal{G}_e$. The electron operator $c_\sigma(i)$ can always be expressed as the sum of holon $\xi_\sigma(i)$ and doublon $\eta_\sigma(i)$,  as detailed in eqn.~[\ref{Eq:cxieta}-\ref{Eq:eta}]. The 2N-component paramagnetic composite operator basis is then defined as 
\begin{align}
\mathbf{\Psi}=\begin{pmatrix}
\xi_{\uparrow}(1), 
...,
\xi_{\uparrow}(N),
\eta_{\uparrow}(1) ,
...,
\eta_{\uparrow}(N)
\end{pmatrix}^T&
\label{Eq:CompOP}
\end{align}

The matrix form of the composite Green's function is written in imaginary time as
\begin{align}
    \mathds{G}(\tau)=-\left\langle T_\tau \left( \mathbf{\Psi}( \tau )  \mathbf{\Psi}^\dagger(0) \right)  \right\rangle 
\end{align}
$\mathds{G}(\tau)$ is computed using equation of motion of $\mathbf{\Psi}$. The current $\mathbf{j}$ for $\mathbf{\Psi}$ is defined as
\begin{align}
&\mathbf{j}(\tau)=-[\mathcal{H},\mathbf{\Psi}](\tau)&
\end{align}
Since the equation of motion does not truncate with the hopping term, we perform a truncation by projecting onto the $\mathbf{\Psi}$ basis. This procedure leads to the definition of three matrices: the $I$, $M$, and $E$ matrices

\begin{align}
\mathds{I} &= \left\langle \{ \Psi(0),\Psi^\dagger(0) \}  \right\rangle     \label{Eq:EOM_IMat}\\
\mathds{M}&= \left\langle \{ \mathbf{j}(0),\Psi^\dagger(0) \}  \right\rangle     \\
 \mathds{E}&= \mathds{M} \mathds{I}^{-1}&
\end{align}

The advanced and retarded composite Green's can be expressed in terms of these matrices as
\begin{align}
    \mathds{G}^{R/A}(\omega)=\left[ (\omega \pm i\eta) \mathds{1} - \mathds{E} \right]^{-1} \mathds{I},
    \label{Eq:Grealomega}
\end{align}
The electronic Green's function can be expressed in terms of the composite Green's function. If we define the electronic basis $\mathbf{c}$ as 
\begin{align}
&\mathbf{c}=\left(c_\uparrow(1),...,c_\uparrow(N)\right)^T&
\end{align}
then $\mathcal{G}_e^{R/A}(\omega)$ is expressed as 
\begin{align}
\left[\mathcal{G}_e^{R/A}(\omega)\right]_{i,j}=&\left[\mathds{G}^{R/A}(\omega)\right]_{i,j}+\left[\mathds{G}^{R/A} (\omega)\right]_{i+N,j} \nonumber \\ &+\left[\mathds{G}^{R/A}(\omega)\right]_{i,j+N}+\left[\mathds{G}^{R/A}(\omega)\right]_{i+N,j+N}
\end{align}\\
In the following, we provide the expression of the components of the current $I$, $M$ and hence $E$ matrices using $S^-(i)=c^\dagger_{\downarrow}(i) c_{\uparrow}(i)$, $S^+(i)=c^\dagger_{\uparrow}(i)c_{\downarrow}(i)$, and  $\Delta(i)=c_{\uparrow}(i)c_{\downarrow}(i)$. For the paramagnetic basis, we can assume $\langle n_{\uparrow}(i) \rangle =\langle n_{\downarrow}(i) \rangle = \frac{n(i)}{2}$.  Using these currents we can calculate the component of the $M$-matrix as follows
\begin{align}
\begin{split}\label{eq:M11}
    \mathds{M}(i,j)=&-\delta_{ij} \left[\mu \left(1-\frac{n(i)}{2}\right) +\sum\limits_l t_{il}e(i,l) \right]\\
         & -t_{ij}\left(1-\frac{n(i)+n(j)}{2}+p(i,j)\right)
\end{split}\\
\begin{split}\label{eq:M12}
   \mathds{M}(i,j+N)=&\delta_{ij}\sum\limits_lt_{il}e(i,l)-t_{ij}\left(\frac{n(j)}{2}-p(i,j)\right)
\end{split}\\
\begin{split}\label{eq:M21}
  \mathds{M}(i+N,j)=&\delta_{ij}\sum\limits_lt_{il}e(i,l)-t_{ij}\left(\frac{n(i)}{2}-p(i,j)\right)
\end{split}\\
\begin{split}\label{eq:M22}
    \mathds{M}(i+N,j+N)=&-\delta_{ij} (\mu-U)\frac{n(i)}{2}\\
         & -\delta_{ij}\sum\limits_l t_{il} e(i,l)-t_{ij}p(i,j)
\end{split}
\end{align}
where we have introduced the following two expectation values on each bond:
\begin{align}
    e(i,j)=\langle\xi_{\downarrow}(j)\xi^\dagger_{\downarrow}(i)\rangle-\langle\eta_{\downarrow}(j)\eta^\dagger_{\downarrow}(i)\rangle 
\end{align}

\begin{align}
p(i,j)=\frac{1}{2} &\left(\langle n_{\downarrow}(i)n_{\downarrow}(j)\rangle +\langle S^-(i)S^+(j) \rangle \nonumber \right.  \\  & \left. -\langle \Delta(i)\Delta^\dagger(j)\rangle + h.c.  \right) 
\end{align}
Finally we can calculate the $I$-matrix, which is diagonal in the $\mathbf{\Psi}$ basis
\begin{align}
&\mathds{I}(i,j)=\delta_{i,j} \left(1-\frac{n(i)}{2}\right),\\
&\mathds{I}(i+N,j+N)=\delta_{i,j} \frac{n(i)}{2}.
\end{align}

The $M$-matrix and the $I$-matrix depend on $(5N+1)$ unknown parameters ($e,p,\mu$). Note that to evaluate the unknown parameter $n(i)$ and $e(i,j)$ we only need the single particle on-site and nearest-neighbor inter-site correlation functions. However, $p(i,j)$ is a two-particle correlation function, and its evaluation relies on a decoupling approximation known as Roth decoupling~\cite{roth1969electron}.
The correlation function is given by for $m$ and $n$ integers from $m=[0,1]=n$ in terms of the retarded and advanced composite Green's function
\begin{align}
    C_{m+1,n+1}(i,j)
    =-&\int \frac{d\omega}{4i\pi} \left[1+\tanh\left(\frac{\beta \omega}{2}\right) \right] \times \nonumber \\ & \Big[\mathds{G}^R(\omega))-{G}^A(\omega))\Big]_{{i+mN},{i+nN}}
    \label{Eq:C0}
\end{align}

The computation of these integrals can be made faster by performing analytically the $\omega$ integration. \\
In terms of the composite correlation function, the electronic density $n$ can be computed as 
\begin{align}
    n(i)=2(1-C_{11}(i,i)-C_{22}(i,i)-C_{12}(i,i)-C_{21}(i,i))
    \label{Eq:ni_re}
\end{align}
The $e(i,j)$ parameter can also be simply expressed in terms of correlation functions
\begin{align}
    e(i,j)=C_{11}(i,j)-C_{22}(i,j).
    \label{Eq:e_re}
\end{align}

Applying the Roth decoupling scheme we calculate the two-point static correlation  $\langle n(i)n(j)\rangle$, $\langle S^-(i)S^+(j)\rangle$ and $\langle \Delta(i)\Delta^\dagger(j)\rangle$ necessary to the computation of $p(i,j)$.
\begin{align}
\langle \Delta(i)\Delta^\dagger(j)\rangle=\dfrac{\rho^\Delta(i,j)}{1+\phi(i)}
\end{align}
\begin{align}
\langle S^-(i)S^+(j)\rangle=\dfrac{-\rho^S(i,j)}{1+\phi(i)}
\end{align}
\begin{align}
\langle n(i)n(j)\rangle=\dfrac{-\rho^S(i,j)}{1-\phi(i)^2}&+\left[ \dfrac{n(j)}{2(1+\phi(i))} \right. \nonumber \\ &\left. \times \left(1-\dfrac{2(C_{11}(i,i)+C_{21}(i,i))}{2-n(i)} \right) \right]
\end{align}
Here we have defined the following variables
\begin{align}
\phi(i)=&\dfrac{2}{n(i)}\big(C_{12}(i,i)+C_{22}(i,i)\big) \nonumber \\ &-\dfrac{2}{2-n(i)}\big(C_{11}(i,i)+C_{21}(i,i)\big)
\end{align}
And defining $C_{mn}(i,j)=C_{mn}$ for brevity, we obtain the following
\begin{align}
\rho^\Delta(i,j)=&\dfrac{2}{2-n(j)}\left(C_{11}+C_{21}\right)\left(C_{22}+C_{21}\right) \nonumber \\ &+\dfrac{2}{n(j)}\left(C_{12}+C_{22}\right)\left(C_{11}+C_{12}\right)
\end{align}
\begin{align}
\rho^S(i,j)=&\dfrac{2}{2-n(j)}\big(C_{11}+C_{12}\big)\big(C_{11}+C_{21}\big) \nonumber \\ 
&+\dfrac{2}{n(j)}\big(C_{22}+C_{12}\big)\big(C_{22}+C_{21}\big)
\end{align}

%\subsection{Uniform systems}
%\label{Appendix:Uniform}
%For translationally invariant systems we perform Fourier transform in k-space, with all $n_i=n$, $p_{ij}=p$ and $e_{ij}=e$. The M-matrix becomes,
%\begin{align}
%&\mathds{M}_{1,1}(\mathbf{k})=-\mu \left(1-\frac{n}{2} \right)-4te- \alpha(\mathbf{k}) \left(1-2n+p\right)&\\
%&\mathds{M}_{1,2}(\mathbf{k})=4te-\alpha(\mathbf{k})\left(\frac{n}{2}-p\right)=\mathds{M}_{2,1}(\mathbf{k})&\\
%&\mathds{M}_{2,2}(\mathbf{k})=-(\mu-U)\frac{n}{2}-4te-\alpha(\mathbf{k})p&
%\end{align}
%where for square lattice with nearest neighbor hopping $\alpha(k_x,k_y)=2t\left(\cos(k_x)+\cos(k_y)\right)$. Whereas $2\times2$ I-matrix is diagonal and independent of $k$ given by
%\begin{align}
%\mathds{I}_D=&\begin{pmatrix}
%1-\frac{n}{2}, 
%\frac{n}{2}
%\end{pmatrix}&
%\end{align}
%Following Ref. one can arrive at the self-consistency equations.

\section{Understanding the Luttinger violation in Hubbard operator methods}
\label{App:LuttingerViolation}
%Luttinger theorem was first proven perturbatively using the Luttinger functional and more recently demonstrated~\cite{oshikawa2000topological} through a topological argument based on $U(1)$ flux insertion. The topological proof relies on the assumption that the system is a Fermi liquid. This assumption is clearly violated in a Mott insulator, so it is expected that Luttinger theorem is violated for Mott insulatros. Several studies suggest that Luttinger's theorem can be preserved at half-filling by incorporating Green's function zeros into the Luttinger count~\cite{dzyaloshinskii2003some,rosch2007breakdown,stanescu2007theory,dave2013absence,skolimowski2022luttinger}. However, from the perspective of the topological argument, zeros do not carry charge and are therefore unaffected by the $U(1)$ flux insertion used in the topological derivation of Luttinger's theorem which establish that Luttinger's theorem is violated at half-filling due to the presence of the Mott insulating state. Away from half-filling, the system becomes metallic, and the violation of Luttinger theorem becomes non trivial. 
\begin{figure}[h!]
\includegraphics[width=0.49\textwidth]{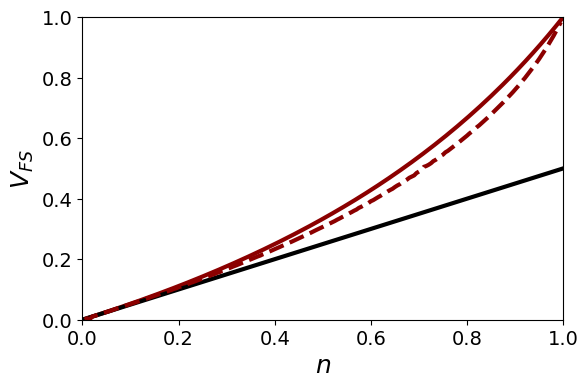}
\caption[0.5\textwidth]{Fermi surface volume as a function of $n$ for a Fermi liquid respecting Luttinger's theorem (thick black line) and for a correlated metal with non-fermionic low-energy excitations, obtained using the Hubbard operator method at finite $U=30t$ (dashed red line), and in the limit of $U\rightarrow\infty$ (thick red line).}
\label{fig:Appfig1}
\end{figure}

In Fig.~\ref{fig:Appfig1}, we demonstrate that the Hubbard operator formalism violates Luttinger's theorem in the doped regime~\cite{osborne2021broken,haurie2024bands}. This violation arises directly from the non-fermionic character of the elementary excitations in a Mott insulator.

To illustrate this, we employ the simplified Hubbard-I approximation~\cite{hubbard1963electron}, which shares the essential feature of decomposing electrons into holon and doublon excitations, similar to the composite operator method, but neglects intersite interactions. The Hubbard-I approach can be expressed using the $E$ and $I$ matrices as:
\begin{align}
\mathds{E}(\mathbf{k}) &= \begin{pmatrix}
(1 - n/2)\epsilon_\mathbf{k} - \mu & (1 - n/2)\epsilon_\mathbf{k} \\
(n/2)\epsilon_\mathbf{k} & (n/2)\epsilon_\mathbf{k} - \mu + U
\end{pmatrix}, \\
\mathds{I} &= \begin{pmatrix}
1 - n/2 & 0 \\
0 & n/2
\end{pmatrix},
\end{align}
where $\epsilon_\mathbf{k}$ is the bare single-particle dispersion. The eigenvalues of $\mathds{E}$ determine the poles of the Green's function, while $\mathds{I}$ encodes the spectral weights of the holon and doublon modes.

The electronic density is fixed by the chemical potential, determined self-consistently from:
\begin{align}
n(i) = &2 - 2\left(\langle \xi(i)\xi^\dagger(i) \rangle + \langle \xi(i)\eta^\dagger(i) \rangle  \nonumber \right. \\ &\left. + \langle \eta(i)\xi^\dagger(i) \rangle + \langle \eta(i)\eta^\dagger(i) \rangle \right).
\end{align}

Apart from the renormalization due to $\mathds{I}$, the formalism closely resembles a two-band non-interacting system, with the non-Hermitian $\mathds{E}$ matrix playing the role of the Hamiltonian. The composite Green's function $\mathds{G}$ and the Green's function of a two-orbital non-interacting system $G_{\text{NI}}$ are defined as:
\begin{align}
\mathds{G}(\omega) &= \left( \omega \mathds{1} - \mathds{E} \right)^{-1} \mathds{I}, \\
\mathcal{G}_{\text{NI}}(\omega) &= \left( \omega \mathds{1} - \mathcal{H} \right)^{-1},
\end{align}
emphasizing the impact of the $\mathds{I}$ matrix. If $\mathds{I} = \mathds{1}$, the model reduces to a non-interacting system that trivially satisfies the Luttinger theorem. However, due to the constraint $c(i) = \xi(i) + \eta(i)$, the weights of holon and doublon sectors satisfy:
\begin{align}
\left[\mathds{I}\right]_{\xi\xi} + \left[\mathds{I}\right]_{\eta\eta} = 1,
\end{align}
with each component being positively defined. Thus, $\mathds{I} = \mathds{1}$ is never realized.

In the hole-doped regime, the relevant spectral weight is that of the holons, where $\left[\mathds{I}\right]_{\xi\xi} < 1$. This reduced weight at each $\mathbf{k}$-point inside the Fermi surface implies an enlarged Fermi volume that counts more than $n$ electrons, violating the standard Luttinger theorem. However, the modified Luttinger count can be restored by scaling the holon Fermi volume $V^\xi_{FS}$ by the holon weight:
\begin{align}
\frac{V^\xi_{FS}}{(2\pi)^d} = \frac{2n}{2 - n} + \mathcal{O}(\langle \xi(i)\eta^\dagger(i) \rangle) \mod 2\pi.
\label{Eq:Lutt}
\end{align}
Equation~\eqref{Eq:Lutt} thus represents a generalized version of Luttinger's theorem for a doped Mott insulator with non-fermionic quasiparticles.

Finally, the non-zero overlap $\langle \xi(i)\eta^\dagger(i) \rangle \ne 0$ introduces a weak Pauli-violation correction. However, in the large-$U$ limit, this term vanishes ($\langle \xi(i)\eta^\dagger(i) \rangle \approx 0$), and hence cannot explain the significant Luttinger theorem violation observed in Fig.~\ref{fig:Appfig1}.

\end{document}